# Analysis of Bacteriostatic Effect of Chinese Herbal Medicine Against *E.coli*

Li Ma[1], Shuangjie Chen[1], Yongguang Yang[1]

**Abstract**

To analyze the bacteriostatic effect of Chinese traditional herbal medicines on E. coli, total 35 different preparations (decoction, volatile oil and distillate) of Chinese traditional herbal medicines were tested using plate culture method. The results showed that 18 preparations of traditional Chinese herbal medicines have different inhibition effect on E. coli in vitro. The results also revealed that different process and combination affect the bacteriostatic effect and different medicines could be used in singles or combined to treat E.coli disease.

[1] Division of Human Genetics, Cincinnati Children's Hospital Medical Center, Cincinnati, Ohio, United States
[2] Department of Cancer Biology, University of Cincinnati, Cincinnati, Ohio, United States
Email: yangy9@mail.uc.edu

## Introduction

Veterinary medicine has a long history and rich application experience, which played an important role in protecting the health of human as well livestock before the appearance of antibiotics[1-4]. With the development of science and technology, due to its advantage in convenience, antibiotics have gradually taken over the role of traditional Chinese herbal medicine in clinic[5-7]. However, due to the widespread use of antibiotics, the treatment resistance of pathogens to antibiotics has been increasing[8-11]. Especially, the problem caused by antibiotic residues in meat products has been drawing the attention for the whole society in recent years [12-14]. E. coli has multi-serotypes with strong variation, which are easily to generate resistance to various antibiotics [15, 16]. Nowadays, increasing attention has been given to tradition Chinese herbal medicine due to its potential as an alternative to antibiotics to solve the treatment resistance problems related to antibiotics[17-19]. To further test the possibility of tradition Chinese herbal medicine in practical use, different kinds and concentrations of tradition Chinese herbal medicine were used in this study to evaluate their antibacterial effects by using plate dilution and filter paper sheet methods in order to screen out the formulations that can be put into clinical application and promote the production of animal husbandry industry.

## Materials and Methods

### 1. Materials

Test strain of E.coli was isolated from broiler chickens. Regular Nutrition Agar and MacConkey Agar were prepared following general protocol and used for E.coli propagation. Preparation and decoction of the following single traditional Chinese herbal medicines were performed in the traditional Chinese herbal medicine lab: Honeysuckle, Radix Saposhnikoviae, Radix, Angelica, forsythia, Almonds, Magnolia, Cyrtomium fortune, Notopterygium root, Catnip, Agastache. The following compound herbal medicines were used in this research: decoction of Yinqiao and Shegan, decoction of Maxinshefang , decoction of  Qianghelianfang, decoction of Xinzhijing, distillate of Yinhua, distillate of Xinzhijing, distillate of Qianghuolianfang and distillate of Maxingshefang.

**1. Methods**

Plate dilution and filter paper tablet methods were employed to test the sensitivity of E. coli to traditional Chinese medicine (single and compound) in *vitro*. For the plated dilution method, 1ml of the original medicine liquid was pipetted into the first test tube; 0.5 ml of the original solution and 0.5 ml of saline were added into the second tube. The double dilution method was followed until the fourth test tube; the fifth test tube was set up as medicine-free negative control. To prepare plate for the test, dump the above liquid into a sterile plate, then added 60 °C sterile common nutrient agar or McKenzie agar quickly and mixed evenly, made of flat. Finally, spread the bacteria evenly onto the agar plate using a sterile stick; put the plates into 37 ℃ incubator for 24 h before monitoring the results.

For the filter paper tablet method, sterile filter paper with a diameter of 0. 6 cm were immersed into the original solution of traditional Chinese herbal medicines for 40, take them out and spare for future use. Apply the medicine immersed tablets onto the agar plates that coated with test E.coli, total 6 tablets for each medicine. In the prepared plate, the sterile operation placed the sensitive drug sheet, each Drug paste 6 tablets. Put the the into 37 ℃ incubator for 24 and evaluate the efficiency of bacteria inhibition by measuring the average diameter of the inhibition zone in each group.

# Results

## 1. Anti-bacterial efficiency in plate-dilution group

The results showed that under low concentration, decoction of Yinqiao and Shegan, decoction of Maxinshefang , decoction of Qianghelianfang, decoction of Xinzhijing, distillate of Yinhua all showed antibacterial effect at different degrees. Decoction of Lianqian, decoction of Xinyi, decoction of Shegan, decoction of Qianghuo, decoction of Huoxiang and distilled oil of Liaoqian all showed significant antibacterial effect at high concentration. Those medicines that did not show clear antibacterial effect under low concentration included distilled oil of Yinhua, distillate of Xinzhijing, distillate of Yinshe, distillate of Qianghuolianfang, distilled oil of Baizhi, decoction of Lianqiao and decoction of Xingren. Those medicines that did not show clear antibacterial effect even in high concentration included distillate of Maxingshefang, decoction of Jingjie and decoction of Baizhi.

## 2. Anti-bacterial result in filter paper tablet group

It was shown in Table 1 that single traditional Chinese herbal medicine distillate of Yinhua, distillate of Guanzhong, decoction of Lianqiao, decoction of Shegan and decoction of Huoxiang all have significant antibacterial effect. However, distilled oil of Fangfeng, decoction of Xinyi decoction of Banlangen and decoction of Qianghuo has the inferior antibacterial effect. In addition, four tradition Chinese herbal medicines have clearly antibacterial effect, those included decoction of Xinzhijing compound, decoction of Qianghuolianfang, decoction of Maxingshefang and decoction of Yinshe.

Table 1  The Antibacterial effect of Traditional Chinese Herbal Medicines on E.coli

| Traditional Chinese Herbal Medicine | Diameter of Inhibition loop (cm) | Average Diameter (cm) | Inhibition Efficiency |
|---|---|---|---|
| Distilled oil of Fangfeng | 0. 75 0. 75 0. 70<br>0. 70 0. 80 0. 80 | 0.75 | + |
| Distillate of Yinhua | 0. 90 0. 95 0. 90<br>0. 90 0. 90 0. 90 | 0.91 | + + |
| Decoction of Banlangen | 0. 80 0. 80 0. 80<br>0. 80 0. 75 0. 65 | 0.77 | + |
| Decoction of Guanzhong | 0. 85 0. 85 0. 80<br>0. 80 0. 90 0. 90 | 0.85 | + + |
| Decoction of Lianqiao | 0. 80 0. 80 0. 80<br>0. 80 0. 85 0. 90 | 0.83 | + + |
| Decoction of Shegan | 0. 80 0. 80 0. 80<br>0. 80 0. 80 0. 80 | 0.80 | + |
| Decoction of Xinyi | 0. 70 0. 70 0. 80<br>0. 85 0. 85 0. 85 | 0.79 | + |
| Decoction of Qianghuo | 0. 70 0. 70 0. 70<br>0. 80 0. 85 0. 90 | 0.78 | + + |
| Decoction of Huoxiang | 0. 80 0. 80 0. 80<br>0. 85 0. 85 0. 90 | 0.83 | + + |
| Distilled oil of Lianqiao | 0. 90 0. 90 1. 00<br>1. 00 1. 10 0. 95 | 0.96 | + + |
| Decoction of Xinzhijing | 0. 90 0. 90 0. 90<br>0. 80 0. 95 0. 95 | 0.90 | + |
| Decoction of Qianghuolianfang | 0. 80 0. 80 0. 80<br>0. 80 0. 80 0. 80 | 0.80 | + + |
| Decoction o Maxingshefang | 0. 70 0. 80 0. 85<br>0. 85 0. 85 0. 85 | 0.82 | + + |
| Decoction of Yinxing | 0. 80 0. 80 0. 80<br>0. 80 0. 85 0. 90 | 0.82 | + + |

Note: It has strong inhibition effect and indicated as "+++" when the diameter of the inhibition zone was larger than 1.0 cm; "++" indicates the diameter of the inhibition zone was between

0.8-1.0 cm; and "+" indicates the weak inhibition effect when the diameter of the inhibition zone was smaller than 0.8 cm.

**Conclusion and Discussion**

Total 14 kind of traditional Chinese herbal medicines (single and compound) screened by plate dilution method showed different degrees of inhibition on E.coli has different degrees of inhibition. Therefore, inhibition zone were observed in all groups when these 14 kinds of traditional Chinese herbal medicines were used for susceptibility test through filter paper tablet method. Because of the difficulty of quantitative analysis of the results, with a subjective assessment frequently introduced and easy to produce errors and affect the accuracy of the results, we did elaborate on the results of plate dilution assays. In this study, plate dilution method was only taken as a qualitative test to screen the antibacterial effect of Chinese herbal medicine; and the filter paper tablet method was taken as a quantitative test to evaluate their antibacterial effect.

It can be concluded from the results that single distillate of Yinhua, Guanzhong, Lianqiao, Shegan and Huoxiang all showed clearly inhibition effect on E.coli. All other single traditional Chinese herbal medicines only have relatively weak inhibition effect on E.coli. Four kind of compound tradition Chinese herbal medicine also demonstrated clearly inhibition effect on E. coli, indicating that the appropriate combination of medicines can improve their antibacterial effect [20-23].

Our results also revealed that compound was much better that single medicine. On one hand, this was probably due to different proportion of the antibacterial agents in the compound[24-26]; on the one hand, the spread patter and method of E.coli on the agar plate will also affect the results, which was consistent with the report from Yuqing Liu et al [27-31].

The inhibitory concentration of tradition Chinese herbal medicine was 1000 times higher than that of antibiotics[32-35]. Thus, it was very difficult to achieve such a high concentration *in vivo* based on the current processing techniques of tradition Chinese herbal medicines [36-39]. It is very difficult to replace antibiotics with only single traditional Chinese herbal medicine[40-44].

In order to reduce drug resistance and antibiotic residues, future study should focus on the combined application of traditional Chinese herbal medicine with antibiotics[45-48].